\documentclass[a4paper,11pt]{article}
\pdfoutput=1 

\usepackage{jinstpub} 
\usepackage{amsthm}
\usepackage{color}
\usepackage{fixltx2e} 
\usepackage{tikz}
\usepackage{subfigure}
\usepackage{txfonts}
\usepackage{amsmath}
\usepackage{natbib}

\title{\boldmath THGEM gain calculations using Garfield++: Solving discrepancies between simulation and experimental data}

\author[1]{C.D.R. Azevedo,\note{Corresponding author.}}
\author{P.M. Correia,}
\author{L.F.N.D. Carramate,}
\author{A.L.M. Silva}
\author{and J.F.C.A. Veloso}

\affiliation{I3N - Physics Department, University of Aveiro\\Campus Universit\'ario de Santiago 3810-193 Aveiro, Portugal}

\emailAdd{cdazevedo@ua.pt}

\abstract{

Discrepancies between the measured and simulated gain in Thick Micropatterned gaseous detectors (MPGD), namely THGEM, have been observed by several groups. In order to simulate the electron avalanches and the gain the community relies on the calculations performed in Garfield++, known to produce differences of 2 orders of magnitude in comparison to the experimental data for thick MPGDs.

In this work, simulations performed for Ne/5\%CH\textsubscript{4}, Ar/5\%CH\textsubscript{4} and
Ar/30\%CO\textsubscript{2} mixtures shows that Garfield++ is able to perfectly describe the experimental data if
Penning effect is included in the simulation. The comparison between the number of excitations which may lead to a
Penning transfer, is shown for THGEM and GEM, explaining the less pronounced gain discrepancies observed in GEM.

}


\begin{document}
\maketitle
\flushbottom

\section{Introduction}
\label{introduction}

The main software tool used for the gas gain simulation in Micropatterned Gaseous Detectors (MPGDs) is
Garfield++ \cite{ref1}, a powerful toolkit for the detailed simulation of gaseous detectors and the physical processes occurring on it. It
uses Monte Carlo microscopic technique to track electrons in gases on a molecular level, retrieving information about each
excited atom: the (\emph{x,y,z}) position, the time of production and the excitation level \cite{ref2}.
The main drawback of using Garfield++ for gain simulations in Thick Gas Electron Multiplier (THGEM) \cite{ref3}
has been the lower gain obtained when compared with the experimental data, reaching up to 2 orders of magnitude
less for measured gains of about 10\textsuperscript{5} (see Figure~\ref{Figure1}). Such discrepancies were
generally assumed to be related with the electric field calculation by the finite-elements software or the charging-up
of the insulator foils. However, recent results in
simulations \cite{ref4,ref5} and experimental data \cite{ref6} of charging-up in GEM and THGEM have showed that the gain variation due to the insulator charging-up is in the order of a few
tens of \%, not explaining the 2 orders of magnitude difference. In this work, due to the long calculation time and
complexity, charging-up effects were not considered. 

One of the input parameters allowed by Garfield++ is the Penning fractions for the gas mixtures, an option which is not
often used by users since the data on Penning fractions is sparse and not reliable. The Penning effect occurs in gas
mixtures when the energy of an excited atom is higher than the ionization potential of the admixture gas (quencher). If
a collision between them occurs, energy transfers are possible resulting in an extra electron \cite{ref7}.
Recently, very detailed and exhaustive studies have been carried out in Penning transfer namely in argon and neon
mixtures \cite{ref7,ref8,ref9}.

Our hypothesis is that the Garfield++ gain deficit is due to the non-inclusion of Penning effect in the calculations
performed by users. In this work we will evaluate Garfield++'s performance in the thick MPGDs gain calculations
by including/excluding the Penning effect and comparing them to experimental data.

\section{Method}
\label{method}

\begin{figure}[htbp]
\centering \includegraphics[width=0.45\textwidth]{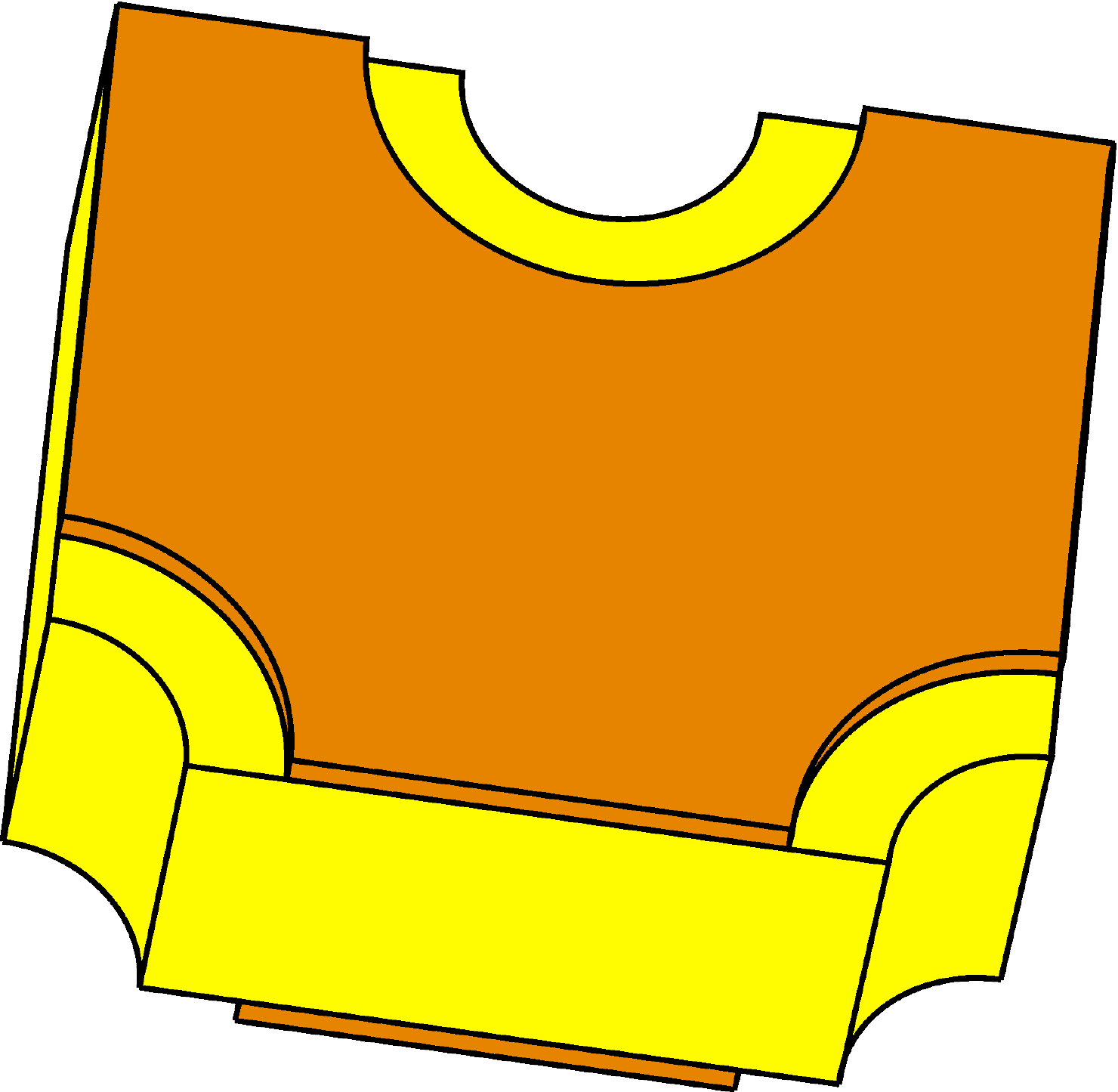}
\caption{Unitary cell used for the THGEM gain calculations. \emph{t}\,=\,0.4\,mm, \emph{p}\,=\,0.7\,mm, \emph{d}\,=\,0.3\,mm, \emph{rim}\,=\,0.1\,mm.}
\label{Figure1}
\end{figure}

In this work a 0.4\,mm thickness (\emph{t}) THGEM was simulated. The hole diameter (\emph{d}) was set to 3\,mm with a 0.1\,mm rim (\emph{rim}) and a 0.7\,mm pitch (\emph{p}) being the unitary cell shown in Figure~\ref{Figure1}. Garfield++'s input electric fields were calculated using Elmer \cite{ref12}, a finite elements electrostatics calculation software, while Gmsh \cite{ref10,ref11}, was used as mesh generator for Elmer.

Calculations in Ne/5\%CH\textsubscript{4}, Ar/5\%CH\textsubscript{4} and Ar/30\%CO\textsubscript{2 }were performed with
and without the inclusion of Penning transfers. Every time the Penning transfers were considered, its rates were set to
$r$\,=\,0.4, $r$\,=\,0.18 \cite{ref8} and $r$\,=\,0.57 \cite{ref7} respectively to the above listed mixtures.

Two other MPGDs were simulated: the Thick WELL (THWELL) \cite{ref13,ref14} (thickness\,=\,0.4\,mm, hole diameter\,=\, 0.5\,mm, rim\,=\,0.1\,mm and pitch\,=\,1.0\,mm) in Ne/5\%CH\textsubscript{4} and a
regular GEM (thickness\,=\,0.05\,mm, biconical hole diameter\,=\,0.07/0.05\,mm, rim\,=\,0.08\,mm and pitch\,=\,0.140\,mm) in
Ar/30\%CO\textsubscript{2}. In this case, the Garfield++'s input electric fields were calculated by ANSYS\textregistered  \,\cite{ref15}.

For all the cases, the drift and induction regions were set to 5\,mm and 2\,mm, with electric fields set to 500V/cm and
2000\,V/cm, respectively.

Gas pressure and temperature were considered to be at 760\,Torr and 20\,\textdegree C. For each condition 1000 events were calculated.
The gain was assumed to be the average number of electrons produced in the gas per drift electron, i.e., the absolute
gain.

\section{Results and discussion}
\label{Results}

\begin{figure}[htbp]
\centering 
\subfigure[]{\label{Figure2}\includegraphics[width=0.45\textwidth]{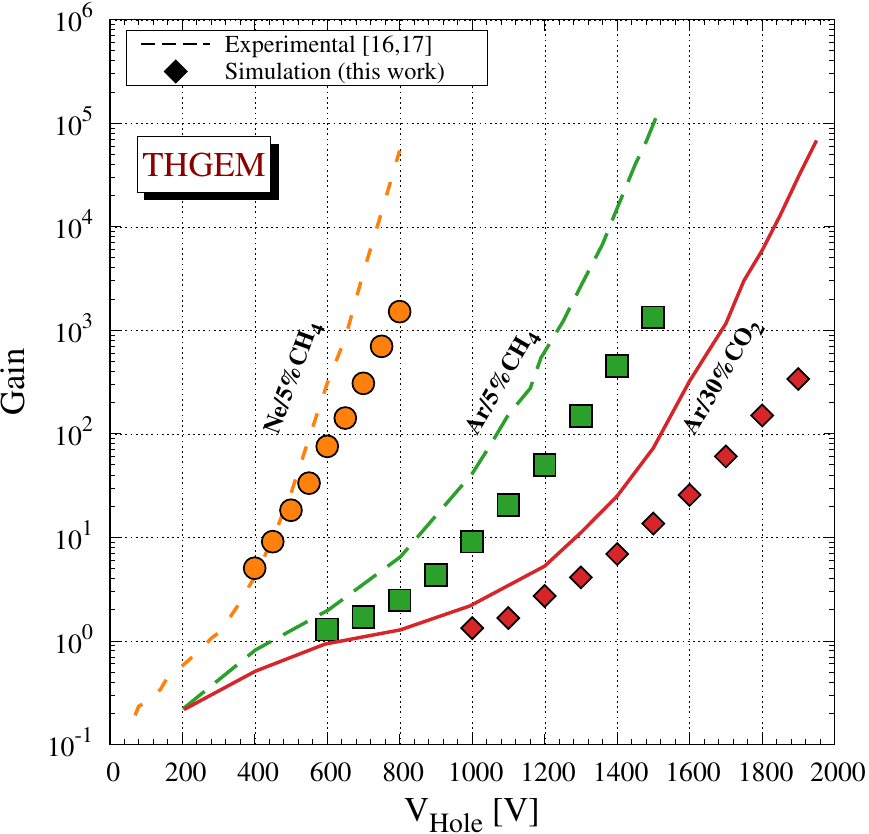}}
\subfigure[]{\label{Figure3}\includegraphics[width=0.45\textwidth]{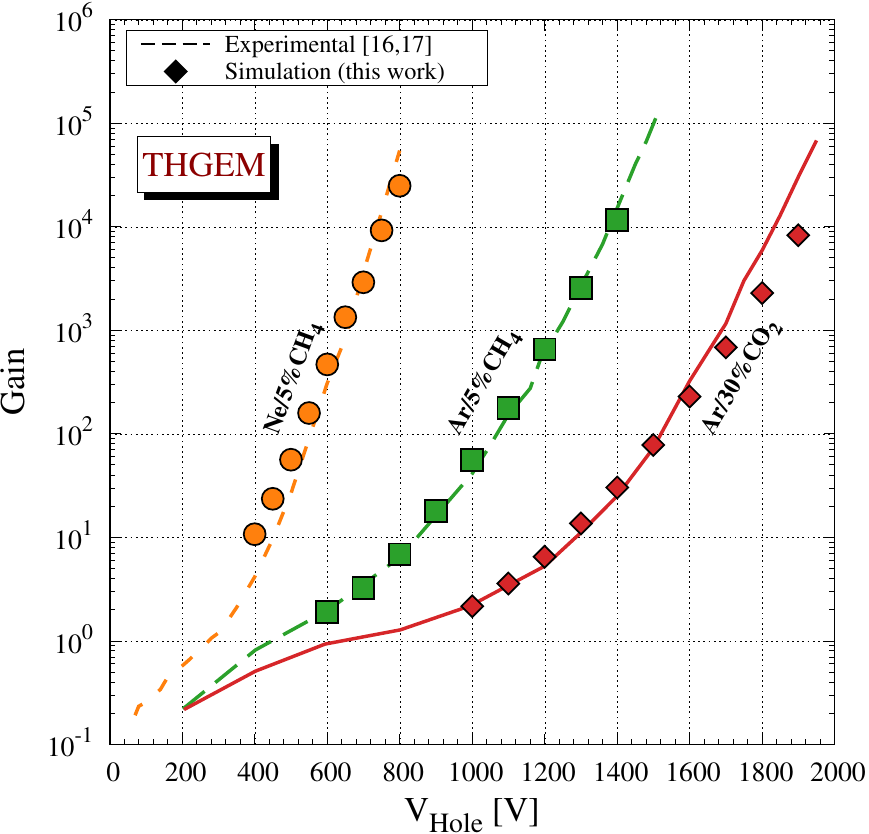}}
\caption{THGEM gain for Ne and Ar based mixtures with CH\textsubscript{4} and CO\textsubscript{2}. Lines are experimental data adapted from \cite{ref16,ref17} while dots are the simulated data. (a) Penning transfers not considered. (b) Penning transfers considered. Error bars are not visible as they  are smaller than the dot size.}
\label{Figure2Tot}
\end{figure}

Figure~\ref{Figure2} shows the results of Garfield++'s gain calculations for a 0.4\,mm THGEM operating in
Ne/5\%CH\textsubscript{4}, Ar/5\%CH\textsubscript{4} and Ar/30\%CO\textsubscript{2} when Penning transfers are
not considered. For comparison, experimental data for the same THGEM geometry is also shown.

The previously described effect, i.e., the lower calculated gain relatively to the experimental data, is clearly observed.
The discrepancies increase with the gain, reaching 2 orders of magnitude when the experimental gains approach
10\textsuperscript{5}.

Calculations for the same gas mixtures were also performed considering Penning transfers. For Ar/5\%CH\textsubscript{4}
a transfer rate $r$\,=\,0.18 taken from \cite{ref8} was used while for Ar/30\%CO\textsubscript{2} a value of $r$\,=\,0.57 was set \cite{ref7}.
The value used for Ne/5\%CH\textsubscript{4} is empirical due to the lack of data and was chosen by performing
preliminary tests. The results are plotted in Figure~\ref{Figure3}.

Comparing the data of Figure~\ref{Figure2} with that of Figure~\ref{Figure3} we can observe an 
improvement of the agreement between experimental and simulated data.  For Ar/5\%CH\textsubscript{4}  we can observe a good match showing that Garfield++ is able to describe the THGEM gain in this mixture. For Ar/30\%CO\textsubscript{2}the simulated gain is in good agreement with the experimental one until it reaches a value of 500.
After that value the simulation results slightly diverge from the experimental data presenting a deviation by a factor
of 2 when the gain reaches 2$\cdot$10\textsuperscript{4}. Such difference is explained in \cite{ref7}
as being due to photon feedback (electrons extracted from the cathode or CO\textsubscript{2} ionization by the
scintillation photons), which was not included in the simulation.
In the case of Ne/5\%CH\textsubscript{4} we can observe a small deviation between the calculations and the experimental data both for low and high gains. The fact that for low gains the calculations present higher values let us consider that maybe, the $r$\,=\,0.4 value could be an overestimation. By other side, when considering the methodology in ref \cite{ref17} we may think that this measurement can have not enough precision (at low gains) for a good estimation on the Penning transfer rates. When looking at high gains we can observe a slightly lower values for calculations relatively to the experimental values. In this case there are two possibilities: or the Penning rate is underestimated, or the most probable in our opinion: photon feedback process should be considered, as described in ref \cite{ref19}.
We wish to claim the reader attention for the fact that $r$\,=\,0.4 is just an estimative. A precise work should be done in order to get a concise value.

\begin{figure}[htbp]
\centering \includegraphics[width=0.45\textwidth]{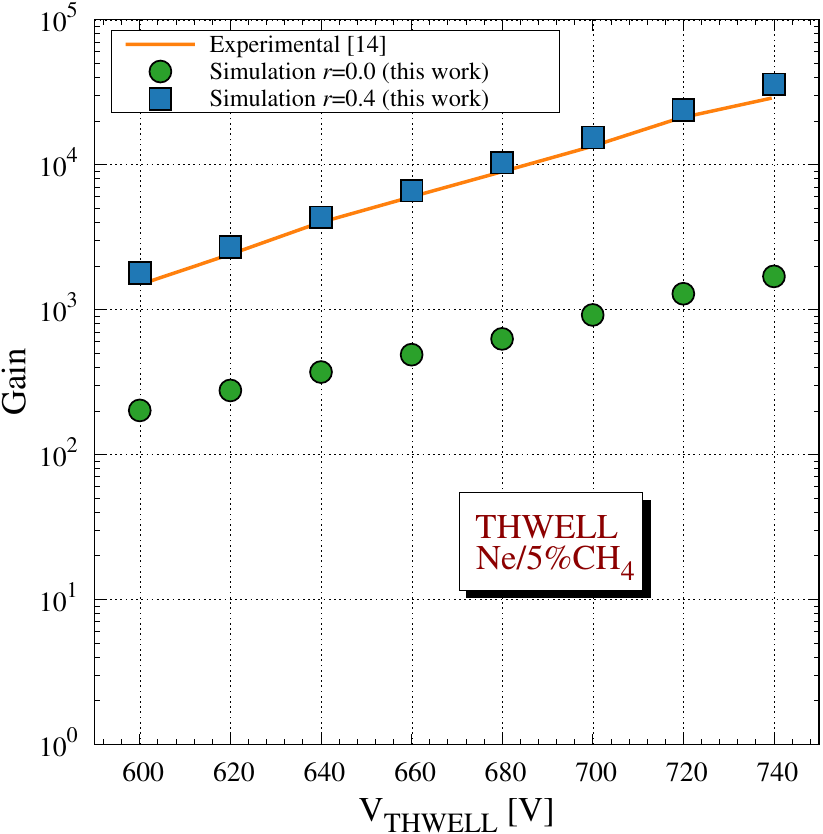}
\caption{THWELL gain in Ne/5\%CH\textsubscript{4}. Line is
experimental data adapted from \cite{ref14}.
Dots are the simulated data. Error bars are not visible as they  are smaller than the dot size.}
\label{Figure4}
\end{figure}

In order to confirm that the gain discrepancy is not caused by the choice of finite element software or multiplier structure, the same calculations were carried out for a THWELL multiplier and using ANSYS\textregistered \,software. The results are presented in Figure~\ref{Figure4} for a Ne/5\%CH\textsubscript{4} mixture with and without considering Penning transfers. The difference between the experimental and calculated data when no Penning transfers are considered is clearly observed. However, when a Penning transfer \ rate of $r$\,=\,0.4 is included, the experimental data is in good agreement with the simulation, showing that Garfield++ is able to perform gain calculations in thick multipliers, accurately modelling the experimental results.

\begin{figure}[htbp]
\centering 
\subfigure[]{\label{Figure5}\includegraphics[width=0.45\textwidth]{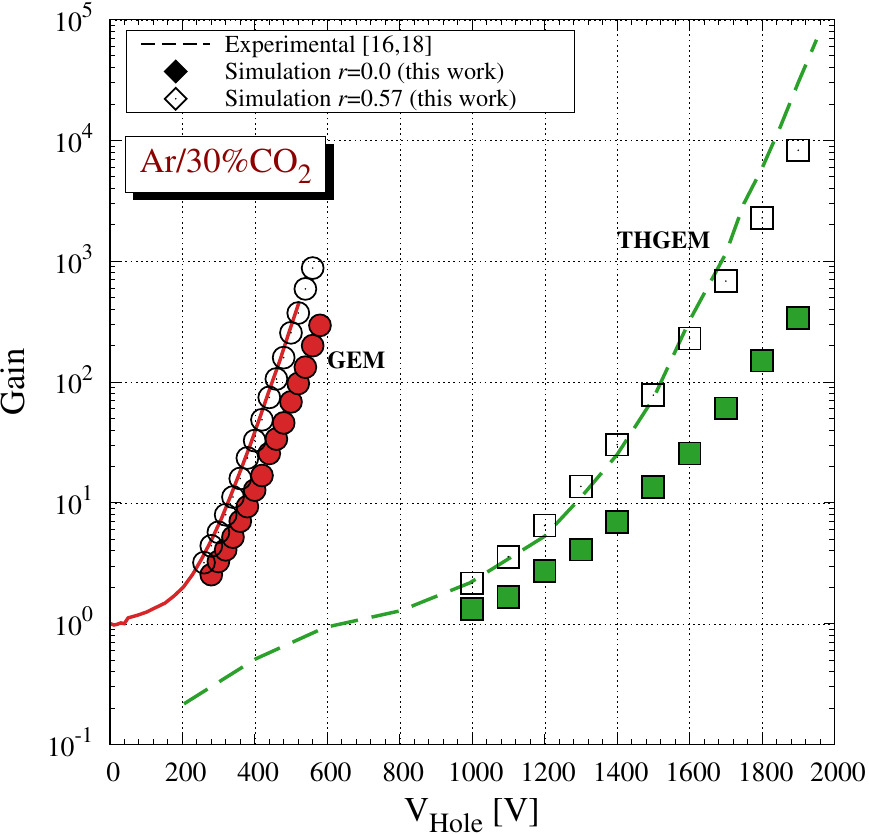}}
\subfigure[]{\label{Figure6}\includegraphics[width=0.45\textwidth]{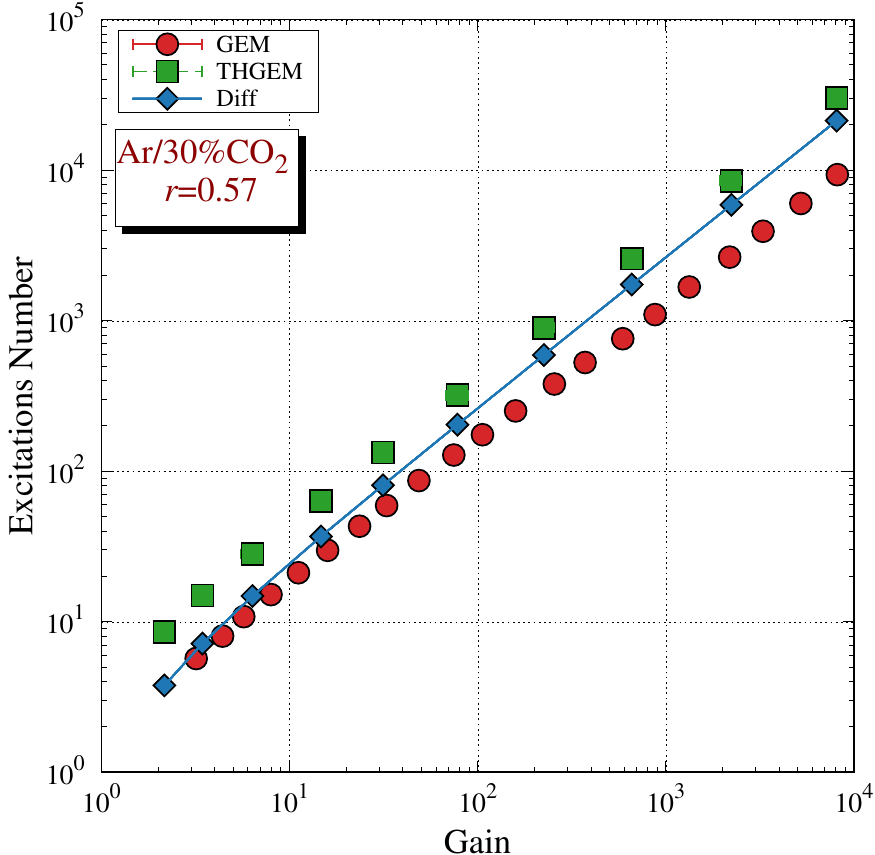}}
\caption{(a) GEM and THGEM gain calculations in Ar/30\%CO\textsubscript{2} including and excluding Penning transfers in the simulation - Lines are experimental data
adapted from \cite{ref16,ref18}. (b) Number of excitation in GEM and THGEM for the previous mixture. The solid line is the difference between the number of excitations in THGEM and GEM. Error bars are not visible as they  are smaller than the dot size.}
\label{Figure5Tot}
\end{figure}

Figure~\ref{Figure5} shows Garfield++ gain calculations for GEM and THGEM in a mixture of
Ar/30\%CO\textsubscript{2}, excluding (closed dots) and including (open dots) the Penning transfers. As can be
observed, in GEMs the Penning effect in the gain discrepancies is less pronounced. In order to understand this effect
we have at look to Garfield++ results keeping in mind that the extra ionization in the gas comes from the excited argon
atoms that transfer the energy to CO\textsubscript{2} ionizing it. The result is shown in Figure~\ref{Figure6},
where we have observed a higher number of excitations in THGEM than in GEM, explaining why the differences in gain is lower in GEM than in THGEM when Penning transfers are not considered.

\section{Conclusions}
\label{Conclusions}

Experimental gain measurements in THGEM and THWELL were successfully modelled by using Garfield++, taking into consideration the Penning transfer rates in the simulation. Significant differences were not observed in the results when using different finite elements software in the calculation of electric fields, thereby excluding this cause for the mismatch between experimental and simulated data.

A Penning transfer rate of $r$\,=\,0.4 was estimated for a Ne/5\%CH\textsubscript{4}, however, for this case a more detailed study should be performed.

We have also observed a stronger effect of the Penning transfers in THGEM than in GEM. This fact is due to the higher ratio of excitations  in THGEM than in GEM. In any case, for correct assessment of the MPGD gains, Penning
transfer rates should always be considered.

\acknowledgments

C.D.R. Azevedo and A.L.M. Silva were supported by Postdoctoral grants from FCT (Lisbon) SFRH/BPD/79163/2011 and
SFRH/BPD/109744/2015, respectively. P.M.M. Correia was supported by the FCT (Lisbon) scholarships BD/52330/2013.

This work was partially support by project PTDC/FIS-NUC/2525/2014 through COMPETE, FEDER and FCT (Lisbon) programs and
pursued within the framework of the CERN RD51 collaboration.

The authors are grateful to I.F. Castro by his work on the language review of this paper.


\end{document}